\documentclass[aps,prl,twocolumn,superscriptaddress]{revtex4}%
\usepackage{eurosym}
\usepackage{amsfonts}
\usepackage{amssymb}
\usepackage{amsmath}
\usepackage{graphicx}
\usepackage{xcolor}%
\providecommand{\U}[1]{\protect\rule{.1in}{.1in}}
\providecommand{\U}[1]{\protect\rule{.1in}{.1in}}
\def\HG#1 {\emph{\color{blue}#1}}
\begin{document}
\title{Structural origin of the anomalous temperature dependence of the local
magnetic moments in the CaFe$_{2}$As$_{2}$ family of materials}
\author{L. Ortenzi}
\affiliation{Institute for Complex Systems (ISC), CNR, U.O.S. Sapienza, v. dei Taurini 19,
00185 Rome, Italy}
\affiliation{Max-Planck-Institut f\"{u}r Festk\"{o}rperforschung,
Heisenbergstra$\mathrm{\beta}$e 1, D-70569 Stuttgart, Germany}
\author{H. Gretarsson}
\affiliation{Max-Planck-Institut f\"{u}r Festk\"{o}rperforschung,
Heisenbergstra$\mathrm{\beta}$e 1, D-70569 Stuttgart, Germany}
\affiliation{Department of Physics, University of Toronto, 60 St.~George St., Toronto,
Ontario, M5S 1A7, Canada}

\author{S.~Kasahara}
\author{Y. Matsuda}
\affiliation{Department of Physics, Kyoto University, Kyoto 606-8502, Japan}

\author{T. Shibauchi}
\affiliation{Department of Advanced Materials Science, The University of Tokyo, Japan}
\affiliation{Department of Physics, Kyoto University, Kyoto 606-8502, Japan}

\author{K. D. Finkelstein}
\affiliation{Cornell High Energy Synchrotron Source, Cornell University, Ithaca, New York
14853, USA}
\author{W.~Wu}
\author{S.~R.~Julian}
\author{Young-June~Kim}
\affiliation{Department of Physics, University of Toronto, 60 St.~George St., Toronto,
Ontario, M5S 1A7, Canada}
\author{I. I. Mazin}
\affiliation{code 6390, Naval Research Laboratory, 4555 Overlook Avenue SW, Washington, DC
20375, USA}
\author{L. Boeri}
\affiliation{Institute for Theoretical and Computational Physics, TU Graz, Petersgasse 16,
8010 Graz, Austria.}
\date{\today }

\begin{abstract}
We report a combination of Fe K$\beta$ x-ray emission spectroscopy and
$ab$-intio calculations to investigate the correlation between structural and
magnetic degrees of freedom in CaFe$_{2}$(As$_{1-x}$P$_{x} $)$_{2}$. The
puzzling temperature behavior of the local moment found in rare earth-doped
CaFe$_{2}$As$_{2}$ [\textit{H. Gretarsson, et al.,
Phys. Rev. Lett. {\bf 110}, 047003 (2013)}] is also observed in CaFe$_{2}%
$(As$_{1-x}$P$_{x}$)$_{2}$. We explain this phenomenon based on
first-principles calculations with scaled magnetic interaction. One scaling
parameter is sufficient to describe quantitatively the magnetic moments in
both CaFe$_{2}$(As$_{1-x}$P$_{x} $)$_{2}$ ($x=0.055$) and Ca$_{0.78}%
$La$_{0.22}$Fe$_{2}$As$_{2}$ at all temperatures. The anomalous growth of the
local moments with increasing temperature can be understood from the observed
large thermal expansion of the $c$-axis lattice parameter combined with strong
magnetoelastic coupling. These effects originate from the strong tendency to
form As-As dimers across the Ca layer in the CaFe$_{2}$As$_{2}$ family of
materials. Our results emphasize the dual local-itinerant character of magnetism in Fe pnictides.

\end{abstract}
\maketitle

It is generally believed that superconductivity in Fe-based materials (FeBS)
is related to magnetic fluctuations.~\cite{treview1,treview2} However, from
the very beginning two schools were formed as regarding the best description
of magnetism in these systems. One described the Fe moments as soft and
itinerant and related magnetic interactions to the Fermi surface
properties.~\cite{Mazin08,Chubukov08,DHLee08} The other assumed localized
moments not unlike those in cuprates that interact via superexchange-type
interactions.~\cite{SiAbrahams,JPHu} At a later stage it was proposed that the
truth is in-between, namely that the moments are substantially localized, and
large, but they are nonetheless soft (can easily change their amplitude, not
just direction), strongly coupled to the lattice (the amplitude strongly depends
on the pnictogen's position), formed by itinerant electrons, and are greatly
reduced by fluctuations.~\cite{MazinJohannesNP,Johannes2009,Parker10,Haule,toschi}

CaFe$_{2}$As$_{2}$ and its derivatives are an odd case among the Fe-based
superconductors (FeBSs). Under pressure the parent compound collapses into a
new structure, with a much shorter $c$-axis lattice parameter and smaller As-As distance.~\cite{Ca122-collapse} 
The crystal structure of CaFe$_{2}$As$_{2}$ at an ambient pressure, like that of all other FeBSs, is very well described by the density functional theory (DFT) calculations, 
\textit{if the moments are allowed to grow to their DFT values,} which can be more than a factor of two larger than the measured \textit{ordered} moments. In those FeBSs in which long-range order is suppressed by means other than pressure, the crystal structure is still
predicted correctly by the magnetic DFT, and not by its nonmagnetic variant.
This strongly suggests that sizeable local moments are present also in FeBSs
which do not show long-range magnetic order. 

Conversely, the high-pressure collapsed tetragonal (cT) structure of
CaFe$_{2}$As$_{2}$ is accurately described by \textit{nonmagnetic} DFT
calculations, { in agreement with the fact that the local moments are quenched
in this phase.~\cite{INS_CT} This magnetic collapse in CaFe$_{2}$As$_{2}$ has
a structural origin, namely the reduced As-As distance across the Ca layer, and
is now fully understood on the basis of DFT
calculations.~\cite{hoffmann,Yildirim,Sanna2012,Valenti} {Experiments also
find} that the magnetic moments in CaFe$_{2}$As$_{2}$ are very sensitive to
this As-As distance.~\cite{preprint-paglione,non-magnetic} }

Besides pressure, structural and magnetic transitions can be modulated through
rare-earth~\cite{rareearth}  or phosphorous~\cite{phase-diagram}  substitutions. 
In this work we will show that in
these compounds not only the ordered moments, but also the local moments are exceptionally soft, and their 
counterintuitive growth with increasing temperature is due to the negative chemical pressure exerted 
upon Fe by the lattice expansion.

Some of us had shown before that the
shape of the Fe K$\beta$ emission line in x-ray emission spectroscopy (XES)
can be used to measure the local moment in FeBSs.~\cite{Hlynur1,Hlynur2} In this paper, we employ this technique to study the temperature dependence of the magnetic moment in CaFe$_{2}$(As$_{1-x}$P$_{x} $)$_{2}$ ($x=0$, $x=0.033$ and $x=0.055$), combined with high-quality X-Ray diffraction data, and show that the anomalous increase of the magnetic moment with temperature correlates with a strong $c$-axis expansion. DFT calculations as a function of the lattice parameter, including an effective fluctuation-induced renormalization of the magnetic
moment,~\cite{Moriya,Ni3Al} quantitatively reproduce the increase of the local
moment in this compound, as well as in the previously measured Ca$_{0.78}$La$_{0.22}$Fe$_{2}$As$_{2}$.~\cite{Hlynur2}
This unambiguously demonstrates that the observed effect is not related to thermal \textit{excitations} but to thermal \textit{expansion}, combined with the unique softness of the Fe moments in CaFe$_{2}$As$_{2}$.

The XES measurements were performed at the Cornell High-Energy Synchrotron
Source (CHESS) on the bending magnet beamline C1. Incident energy of 9 keV 
(1$\%$ bandwidth) was selected from a multilayer monochrometer. To measure the Fe K$\beta$
emission line, we used five Ge(620) 4" spherical analyzers (1 m radius) and a
Vortex detector in quasi-Rowland circle geometry. X-ray diffraction
measurements were performed using a Cu tube source with a Ge(111)
monochromator. Samples were aligned within a four-circle Huber diffractometer.
The temperature dependence of the $c$-axis was determined by monitoring the
(008) Bragg peak. Data was collected on cooling using a closed-cycle
refrigerator. Details of the growth of the single-crystal samples have been
reported before.~\cite{CaFe2As2_growth:2010}

\begin{figure}[th]
\begin{center}
\includegraphics[angle=0, width=1 \columnwidth]{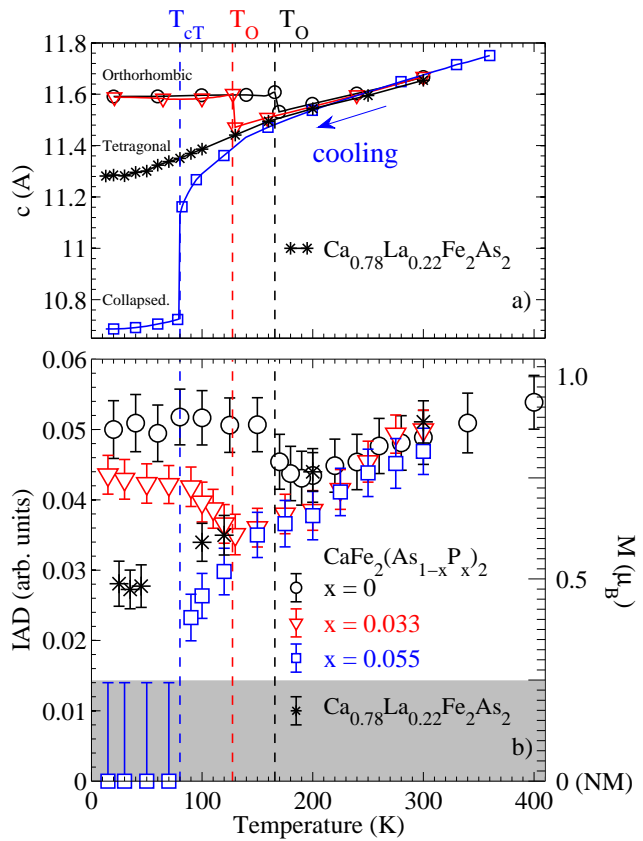}\hfil
\end{center}
\caption{ (Color online) (a) Temperature dependence of the $c$-axis lattice
parameter in CaFe$_{2}$(As$_{1-x}$P$_{x}$)$_{2}$. Both $x=0$ and $x=0.033$
samples go through tetragonal- to orthorhomic-phase transition
(T$_{\mathrm{{O}}}$) around 165 K and 130 K, respectively. In the $x=0.055$ an
abrupt transition into collapsed-tetragonal phase is seen at 80 K
(T$_{\mathrm{{cT}}}$). (b) The temperature dependence of the IAD values for
the same samples derived from their XES spectra. On the right-hand side is the
magnetic moment of Fe ($M$) derived as described in text. Data for Ca$_{0.78}%
$La$_{0.22}$Fe$_{2}$As$_{2}$ -- Ref.~\citenum{Hlynur2} -- are also shown
for comparison.}%
\label{fig01}%
\end{figure}

The ground state of CaFe$_{2}$(As$_{1-x}$P$_{x})_{2}$ has been shown to be
very sensitive to small inhomogeneity in doping and crystal growth conditions;~\cite{phase-diagram, Canfield_QuenchingCa122:2011}  we thus carefully
characterized all our samples using x-ray diffraction. In Fig. \ref{fig01}
(a) the thermal evolution of the $c$-axis lattice constant in CaFe$_{2}%
$(As$_{1-x}$P$_{x})_{2}$ can be seen. Both the parent compound and the
$x=0.033$ sample exhibit a first-order tetragonal(T)- to orthorhomic(O)-phase
transition (T$_{\mathrm{{O}}}$) around 165 K and 130 K, respectively. Upon
increasing $x$ further, T$_{\mathrm{{O}}}$ gets suppressed and eventually
superconductivity appears for $x\sim0.04$. \cite{phase-diagram} At $x=0.055$ superconductivity
disappears and a sharp transition into the collapsed-tetragonal(cT) phase is
seen at 80 K (T$_{\mathrm{{cT}}}$). For comparison we also include data on
Ca$_{0.78}$La$_{0.22}$Fe$_{2}$As$_{2}$ from Ref.~\citenum{Hlynur2}. This
compound stays in the T-phase over the entire temperature range and thus
provides ``a bridge'' between the $x=0.033$ and $x=0.055$ samples.

Fig. \ref{fig01} (a) already reveals the unique feature of CaFe$_{2}%
$As$_{2}$ materials, an enormous lattice constant change with
temperature. Even for the samples that do not exhibit the cT-phase transition
(e.g. $x=0.033$), the c-axis contraction ($\sim$100$\times10^{-5}$
$\mathrm{\mathring{A}/K}$ ) in the T-phase is already  three times larger than
what we have observed in the related compound BaFe$_{2}$As$_{2}$ ($\sim
$35$\times10^{-5}$ $\mathrm{\mathring{A}/K}$).~\cite{Hlynur3}  Interestingly, this rate of
contraction is not seen in the magnetically ordered O-phase, emphasizing the
important role of magnetism in these materials.

In order to investigate if this large change in the $c$-axis lattice parameter
is coupled to the magnetic moment of Fe, we carried out Fe K$\beta$ $(3p
\rightarrow1s)$ XES experiment. By applying the integrated 
absolute difference (IAD) analysis~\cite{Vanko2006} on the shape of the emission line information on the size
of the Fe magnetic moment can be obtained (see Supplemental Material for details).
As in our previous work,~\cite{Hlynur1,Hlynur2} the reference spectra for low-spin and
high-spin were FeCrAs and $\mathrm{K_{2}Fe_{4}Se_{5}}$, respectively. In
Fig.~\ref{fig01} (b) the temperature dependence of the IAD values for all
samples can be seen. Since the IAD is found to be linearly proportional to the
magnetic moment of Fe ($M$)~\cite{VankoPRB2006} we can add to the plot a
right-hand axis showing the local moment scale, as determined from the IAD
values of FeCrAs and $\mathrm{K_{2}Fe_{4}Se_{5}}$. The detection limit of the
IAD technique is shown as a shaded area. \cite{Hlynur2} By comparing our results on
CaFe$_{2}$(As$_{1-x}$P$_{x})_{2}$ with the one obtained for Ca$_{0.78}%
$La$_{0.22}$Fe$_{2}$As$_{2}$ we notice a similar temperature behavior
of the local moment. At room temperature all samples have local moments around
0.8 $\mathrm{\mu_{B}}$, which upon cooling gradually decreases. The $x=0.055$
sample shows the strongest effect, going from $\approx0.8$ $\mathrm{\mu_{B}}$
at $T = 300$ K, to 0.4 $\mathrm{\mu_{B}}$ at 100 K, and in the cT-phase it
drops to zero. A similar trend  can be seen in the T-phase for both the parent
compound and $x=0.033$, although to a lesser extent. Interestingly, in their
magnetically ordered O-phase an increase in the moment is observed which gradually
becomes temperature independent upon further cooling.

Our observation in Fig.~\ref{fig01} highlights the similarities between the
evolution of the $c$-axis lattice parameter and the measured magnetic moment for
both CaFe$_{2}$(As$_{1-x}$P$_{x})_{2}$ and Ca$_{0.78}$La$_{0.22}$Fe$_{2}%
$As$_{2}$. In particular, both $M$ and $c$ strongly depend on temperature in
the T-phase, but not in the antiferromagnetic O-phase, demonstrating a strong
entanglement between the structural and magnetic degrees of freedom.

The unusual thermal sensitivity of the $c$-axis lattice
parameter in CaFe$_{2}$As$_{2}$ and its derivatives, combined with the well-known sensitivity of the magnetic
moments in Fe pnictides to the Fe-As and As-As distances \cite{YPZin,Mazin:DFTproblem}, hints at the
importance of the lattice geometry and thermal expansion. {In order to
investigate a possible \textit{structural} origin of the anomalous behavior
of the magnetic moment}, we performed first-principles calculations {of the
magnetic moment} as a function of the experimental, temperature-dependent
lattice parameters for both La-doped and P-doped ($x=0.055$) samples from
Refs.~\citenum{phase-diagram,rareearth}.~\cite{wien2k,functional,structure}

{In order to correct the well-known tendency of standard local (spin) density approximation - L(S)DA - and generalized gradient approximation (GGA) exchange
and correlation ($xc$) functionals to overestimate the size of the magnetic
moment in Fe pnictides, we used the Reduced Stoner Theory (RST) introduced by
some of us in Ref.~\citenum{Ni3Al}. In practice, we introduced a scaling
parameter $0<s<1$ for the spin-dependent part of the exchange and correlation
functional, which accounts for the reduction of the magnetic moment due to
spin fluctuations, in the spirit of Moriya's self-consistent theory.~\cite{Moriya} The value of $s$ in a given material indicates the importance of spin fluctuations: $s=1$ reproduces the standard exchange and correlation functional (i.e. no additional fluctuation effects are included), while for $s=0$ fluctuations are so strong that magnetism is entirely suppressed.~\cite{Ni3Al} } Importantly, while the original Moriya's theory
 only deals with the ordered moment, by integrating
over all fluctuations with all frequencies, one expects the same approach, but
with reduced renormalization (smaller $1-s$) to work also for the local
moments. Obviously, the scaling parameter in this case should depend on the
time scale of the experiment probing the local moments; since in our
experiment the time scale is independent of doping and temperature, we should
be able, if this theory is correct, to explain the experiment using the same
$s$ for all samples and temperatures.

With this in mind, we have fixed $s$ so as to reproduce the magnetic moment at
the highest temperature (T = 300 K) for the La-doped sample, and found that without
changing $s$ the full temperature dependence of
the magnetic moment for all measured samples could be reproduced.

\begin{figure}[t]
\begin{center}
\includegraphics[trim=0.0cm 0cm -1.5cm 0cm, angle=0, width=.95 \columnwidth]{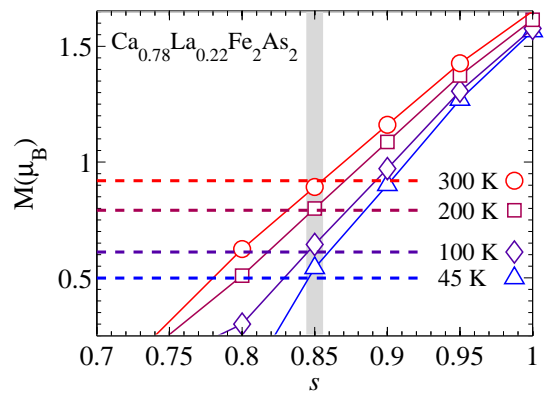}
\end{center}
\caption{{{(Color online) Magnetic moment $M$ of Ca$_{0.78}$La$_{0.22}$%
Fe$_{2}$As$_{2}$ as a function of the DFT-RST scaling parameter $s$ for
different temperatures (solid lines). The horizontal dashed lines show the
experimental values of the magnetic moment for different temperatures taken
from Ref.~\citenum{Hlynur2}. Best agreement with experimental values is
obtained using $s=0.85$ (gray vertical line). }}}%
\label{fig03}%
\end{figure}

{In Fig.~\ref{fig03} we show the scaling effect for Ca$_{0.78}$%
La$_{0.22}$Fe$_{2}$As$_{2}$.  For $s=1$,
corresponding to standard LSDA calculations, the magnetic moment is almost
temperature-independent, but for smaller $s$ the moment acquires a stronger
T-dependence, until magnetism is entirely suppressed below $s\approx0.75$. The
value of $s=0.85$ reproduces experimental data at all temperatures nearly
perfectly. This is even more clearly seen in Fig.~\ref{fig04} (a), where the magnetic moments for $s=0.85$ and $s=1$ are plotted as a function of temperature.

\begin{figure}[t]
\centering
\includegraphics[trim=0cm 0cm -1cm 0cm, angle=0, width=.95 \columnwidth]{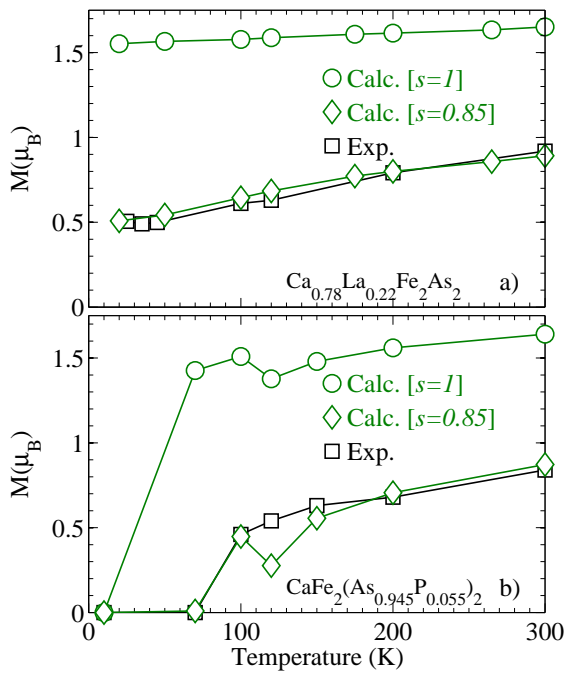}
\caption{(Color online) Magnetic moment as a function of temperature in the
DFT-RST calculations using $s=0.85$ (diamonds) compared with the experimental
data (squares) of (a) Ca$_{0.78}$La$_{0.22}$Fe$_{2}$As$_{2}$
(Ref.~\citenum{Hlynur2}) and (b) CaFe$_{2}$As$_{0.945}$P$_{0.055}$. The
hollow circles indicate conventional LSDA results (equivalent to DFT-RST with
$s=1.00$).}%
\label{fig04}%
\end{figure}

Fig.~\ref{fig04} (b) shows the same calculations for
the P-doped samples at $x=0.055$ (blue squares in Fig.~\ref{fig01} (b)).
Experimentally, this sample turns into a non-magnetic collapsed tetragonal
phase below T$_{\rm cT}$ $=$ 80 K; the complete set of structural parameters and
internal coordinate of the As is published in Ref.~\citenum{phase-diagram}.
As can be seen in Fig.~\ref{fig04} (b), the standard DFT overestimates
$M$ and underestimates its temperature dependence, although it correctly
converges to a nonmagnetic ground state in the cT-phase ($T=10$ K). The same
$s=0.85$ as used for the La-doped compound reproduces remarkably well the
experimental behavior.~\cite{note2} Interestingly, we had not been
able to describe with a single $s$ the behavior of the
ordered moment near the doping-induced quantum critical point (QCP) in
BaFe$_{2}$As$_{2}$.~\cite{unpublished} This is understandable, because (i) a QCP generates much stronger fluctuations than a first-order structural collapse and (ii) more fluctuations contribute to renormalization of the ordered moment than to the local one.

In DFT, the origin of the magnetic collapse of the 122 structure is a subtle
balance between the energy gain induced by lattice collapse, discussed in a
seminal paper by Hoffmann and Zheng,~\cite{hoffmann} and the magnetic energy
gained in the uncollapsed phase. The magnetic stabilization energy in Fe
pnictides is discussed in detail in Ref.~\citenum{lboka}, using a simplified
non-self-consistent version of reduced Stoner theory. This quantity depends in
a non-trivial way on the value of the magnetic moment and on the details of
the electronic structure, and a discussion is beyond the scope of the present
paper. Two observations are however in place: (a) an anti-ferromagnetic (AFM)
ground state can be stabilized only if $d_{xy}$ electronic states are present
at the Fermi level in the non-magnetic DFT band structure;~\cite{Valenti} (b) the transition from the uncollapsed to the collapsed tetragonal phase results in a relative shift of the $d_{xy}$ states with respect to other $d$ states, but the change in occupation is so small ($\Delta n_{xy}\simeq0.1$ $e$/unit cell)
that it does not make sense to talk of a spin-state transition.

A pronounced temperature-dependence of a magnetic moment is often associated
with some thermal excitations. With this in mind,  the experimental data on
Ca$_{0.78}$La$_{0.22}$Fe$_{2}$As$_{2}$, were fitted in Ref.~\citenum{Hlynur2}
with a model assuming thermally excited low-high spin transitions, similar to
the strongly correlated insulator LaCoO$_{3}$~\cite{Saitoh1997}. This model
implies narrow Fe $d$ levels that could form two nearly degenerate magnetic
states,  whose relative occupation changes with temperature. 
An essential part of this model is an assumption that some or all electrons form 
localized narrow states, separated by a crystal field comparable with the Hund's
rule coupling, so that localized electrons on a site can form two nearly degenerate
states, a singlet and a triplet. An alternative version of the same model
instead of using temperature to mix these two states utilizes an empirical
magnetoelastic coupling. A proper tuning
of the model ensured a good description of the experimental
data.~\cite{Hlynur2,giniyat} However, the fundamental assumption of this model
does not seem to be applicable to CaFe$_{2}$As$_{2}$. 
First, as opposed to LaCoO$_{3}$, in CaFe$_{2}$As$_{2}$ the bandwidths for
all orbitals are large, and, most
importantly, all the crystal field splittings are much smaller than either
 the Hund's rule coupling $J$ or the bandwidths. Thus, in actuality the materials appear to be in a completely different regime, where the softness of the magnetic moment 
is derived not from an accidental degeneracy of two ionic states ($i.e.$, from
a competition between the Hund's rule and crystal field) but from the 
competition between the Hund's rule and kinetic energy. Our approach is aimed 
at capturing exactly this competition. The fact that only with one parameter we were able to describe two different, albeit similar material, testifies
that this is the correct physics.

In conclusion, in this work we have addressed the issue of the intriguing and
counterintuitive growth of local magnetic moments with temperature in the
CaFe$_{2}$As$_{2}$ family of materials. We have shown that this phenomenon is
present not only in the La-doped system,~\cite{Hlynur2} but also in CaFe$_{2}%
$(As$_{1-x}$P$_{x}$)$_{2}$, suggesting that it is universal for all Ca-based
122 materials, and ruling out explanations requiring fine parameter tuning.
Using density functional theory, we found that in both materials, again
counterintuitively, the observed strong temperature dependence is not related
to any thermally excited process, but is a consequence of the anomalously
large thermal expansion along $c$, combined with strong magnetoelastic
coupling, both due to As-As interaction across the (relatively thin) Ca layer.
Therefore cooling the sample is equivalent to applying pressure on it.
This finding emphasizes the dual character of magnetism in the parent
materials of Fe-based superconductors, where sizable local moments are
nonetheless soft and exhibit many traits of itinerant magnetism.

\acknowledgements{The authors acknowledge useful discussions with G. Khaliullin. IIM acknowledge
support from ONR through NRL basic research program, and from the A.v.
Humboldt foundation. Research at the University of Toronto was supported by the NSERC, CFI, OMRI, and CIfAR.
Use of the Cornell High Energy Synchrotron Source (CHESS) was supported by the National Science Foundation 
and the National Institutes of Health/National Institute of 
General Medical Sciences under NSF award DMR-0936384. LO
acknowledge support from the Deutsche Forschungsgemeinschaft under
Priority Program 1458, grant number Boe/3536-1.}


\begin{thebibliography}{99}                                                                                               %


\bibitem {treview1}P.J. Hirschfeld, M.M. Korshunov, I.I. Mazin
Reports on Progress in Physics \textbf{74}, 124508 (2011).

\bibitem {treview2}A. Chubukov, Ann. Rev. of Cond. Matt. Phys, \textbf{3}, 57 (2012).

\bibitem {Mazin08}I.I. Mazin, D.J. Singh, M.D. Johannes, and M.H. Du, Phys.
Rev. Lett. \textbf{101}, 057003 (2008).

\bibitem {Chubukov08}A. V. Chubukov, D. V. Efremov, and I. Eremin Phys. Rev. B
\textbf{78}, 134512 (2008).

\bibitem {DHLee08}Fa Wang, Hui Zhai, and Dung-Hai Lee, Europhys. Lett.
\textbf{85}, 37005 (2009).

\bibitem {SiAbrahams}Qimiao Si, Elihu Abrahams, Phys. Rev. Lett. \textbf{101},
076401 (2008).

\bibitem {JPHu}Kangjun Seo, A. B. Bernevig and JiangPing Hu Phys. Rev. Lett
\textbf{101}, 206404 (2008).

\bibitem {MazinJohannesNP}I.I. Mazin and M.D. Johannes Nature Physics
\textbf{5}, 141 (2009).

\bibitem {Johannes2009}I.I. Mazin and M.D. Johannes Phys. Rev. B \textbf{79},
220510 (R) (2009).

\bibitem {Parker10}S.J. Moon, J.H. Shin, D. Parker, W.S. Choi, I.I. Mazin,
Y.S. Lee, J.Y. Kim, N.H. Sung, B.K. Cho, S.H. Khim, J.S. Kim, K.H. Kim, and
T.W Noh, B \textbf{81}, 205114 (2010).

\bibitem {Haule}Z. P. Yin, K. Haule, and G. Kotliar, Nat. Mat. \textbf{10},
932 (2011).
\bibitem {toschi}P. Hansmann, R. Arita, A. Toschi, S. Sakai, G. Sangiovanni, and K. Held Phys. Rev. Lett. \textbf{104}, 197002 (2010).
\bibitem {Ca122-collapse}A. Kreyssig, M. A. Green, Y. Lee, G. D. Samolyuk, P.
Zajdel, J. W. Lynn, S. L. Bud'ko, M. S. Torikachvili, N. Ni, S. Nandi, J. B.
Le$\tilde{\text{a}}$o,S. J. Poulton, D. N. Argyriou, B. N. Harmon, R. J.
McQueeney, P. C. Canfield, and A. I. Goldman Phys. Rev. B \textbf{78} 184517 (2009).
\bibitem {INS_CT}Pratt, D. K et al.
Phys. Rev. B \textbf{79}, 060510( R) (2009).
\bibitem {Yildirim}T. Yildirim, Phys. Rev. Lett. \textbf{102}, 037003 (2009).

\bibitem {Sanna2012}A. Sanna, G. Profeta, S. Massidda, and E. K. U. Gross,
Phys. Rev. B \textbf{86}, 014507 (2012).

\bibitem {Valenti}R. S. Dhaka, R. Jiang, S. Ran, S. L. Bud'ko, P. C.
Canfield,M. Tomi$\breve{c}$ and R. Valent$\acute{i}$, Y. Lee, B. N. Harmon and
A. Kaminski, Phys. Rev. B \textbf{89}, 02511(R) (2014).

\bibitem {hoffmann}R. Hoffmann and C. Zheng J. Phys. Chem., \textbf{89} (20),
4175 (1985).

\bibitem {preprint-paglione}J. R. Jeffries, N. P. Butch, M. J. Lipp, J. A.
Bradley, K. Kirshenbaum, S.R. Saha, J. Paglione, C. Kenney-Benson, Y. Xiao, P.
Chow, and W. J. Evans, arXiv 1401.7400 (2014).
\bibitem {non-magnetic}J. H. Soh, G. S. Tucker, D. K. Pratt, D. L. Abernathy,
M. B. Stone, S. Ran, S. L. Bud'ko, P. C. Canfield, A. Kreyssig, R. J.
McQueeney, A. I. Goldman Phys. Rev. Lett. \textbf{111}, 227002 (2013).
\bibitem {rareearth}S. R. Saha, N. P. Butch, T. Drye, J. Magill, S. Ziemak, K.
Kirshenbaum, P. Y. Zavalij, J. W. Lynn, and J. Paglione Phys. Rev. B {\bf 85} 024525 (2012).
\bibitem {phase-diagram}S. Kasahara, T. Shibauchi, K. Hashimoto, Y. Nakai, H.
Ikeda, T. Terashima, and Y. Matsuda Phys. Rev. B \textbf{83} 060505(R) (2011).

\bibitem {Hlynur1}H. Gretarsson, A. Lupascu, Jungho Kim, D. Casa, T. Gog, W.
Wu, S. R. Julian, Z. J. Xu, J. S. Wen, G. D. Gu, R. H. Yuan, Z. G. Chen, N.-L.
Wang, S. Khim, K. H. Kim, M. Ishikado, I. Jarrige, S. Shamoto, J.-H. Chu, I.
R. Fisher, and Young-June Kim Phys. Rev. B \textbf{84}, 100509(R) (2011).

\bibitem {Hlynur2}H. Gretarsson, S. R. Saha, T. Drye, J. Paglione, Jungho Kim,
D. Casa, T. Gog, W. Wu, S. R. Julian, and Young-June Kim, Phys. Rev. Lett.
\textbf{110}, 047003 (2013).
\bibitem {Moriya}T. Moriya, Spin Fluctuations in Itinerant Electron Magnetism,
Springer Series in Solid-State Science (1985).

\bibitem {Ni3Al}L. Ortenzi, I. I. Mazin, P. Blaha, and L. Boeri, Phys. Rev. B
\textbf{86}, 064437 (2012).

\bibitem {CaFe2As2_growth:2010}Kasahara, S. and Shibauchi, T. and Hashimoto,
K. and Ikada, K. and Tonegawa, S. and Okazaki, R. and Shishido, H. and Ikeda,
H. and Takeya, H. and Hirata, K. and Terashima, T. and Matsuda, Y., Phys. Rev.
B 81 184519 (2010).


\bibitem {Canfield_QuenchingCa122:2011}Ran, S. \emph{et~al.}, \newblock Phys.
Rev. B \textbf{83}, 144517 (2011).

\bibitem {Hlynur3}H. Gretarsson et al., \emph{unpublished}.

\bibitem{Vanko2006}
G. Vanko {\em et~al.},
\newblock J. Phys. Chem. B {\bf 110}, 11647 (2006).
\bibitem {VankoPRB2006}G. Vanko, J. P. Rueff, A. Mattila, Z. Nemeth and A.
Shukla, \newblock Phys. Rev. B \textbf{73}, 024424 (2006).
\bibitem{YPZin} Yin, Z. P. and Lebegue, S. and Han, M. J. and Neal, B. P. and Savrasov, S. Y. and Pickett, W. E., Phys.Rev. Lett. \textbf{101}, 047001
(2008).

\bibitem{Mazin:DFTproblem} Mazin, I. I. and Johannes, M. D. and Boeri, L. and Koepernik, K. and Singh, D. J., Phys.Rev. B \textbf{78}, 085104
(2008).



\bibitem {wien2k}http://www.wien2k.at.
\bibitem{LSDA} J.P. Perdew and Y. Wang, Phys.Rev. B \textbf{45}, 13244
(1992).

\bibitem {functional}All calculations were performed with the full-LAPW code
Wien2K~\cite{wien2k}, using a LSDA exchange and correlation functional.~\cite{LSDA} Up to 3374 $\mathbf{k} $ points in the irreducible wedge were used in the self-consistent calculations, corresponding to a $30\times30\times27$ mesh.
\bibitem {PBE}J. P. Perdew, K. Burke, and M. Ernzerhof, Phys. Rev. Lett.
\textbf{78}, 1396 (1997).
\bibitem {structure}For the La-doped compound the position of the As atoms was determined at each temperature with a DFT optimization, based on the standard PBE exchange and correlation functional,~\cite{PBE} and not the fluctuation-corrected DFT described below.


\bibitem {note2}One may note some disagreement at $T=120$ K; however, the
error bar on this particular point is large, since the experimental data for
the internal As coordinates at this temperature have a large uncertainity. In
fact, for this (and only this) point the optimized As coordinate differs
substantially from the reported experimental one.


\bibitem {unpublished}L. Ortenzi, I. I. Mazin, and L. Boeri,
\emph{unpublished}.


\bibitem {lboka}O.K. Andersen, L. Boeri Ann. Phys. \textbf{1}, 8-50 (2011).

\bibitem {Saitoh1997}T. Saitoh \emph{et~al.}, \newblock Phys.\ Rev.\ B
\textbf{55}, 4257 (1997).

\bibitem {giniyat} J. Chaloupka and G. Khaliullin, Phys. Rev. Lett. \textbf{110}
207205 (2013).














%



















%





\end{thebibliography}
\end{document}


\title{Supplemental Material: Structural origin of the anomalous temperature dependence of the local
magnetic moments in the CaFe$_{2}$As$_{2}$ family of materials}

\author{L. Ortenzi}
\affiliation{Institute for Complex Systems (ISC), CNR, U.O.S. Sapienza, v. dei Taurini 19,
00185 Rome, Italy}
\affiliation{Max-Planck-Institut f\"{u}r Festk\"{o}rperforschung,
Heisenbergstra$\mathrm{\beta}$e 1, D-70569 Stuttgart, Germany}
\author{H. Gretarsson}
\affiliation{Max-Planck-Institut f\"{u}r Festk\"{o}rperforschung,
Heisenbergstra$\mathrm{\beta}$e 1, D-70569 Stuttgart, Germany}
\affiliation{Department of Physics, University of Toronto, 60 St.~George St., Toronto,
Ontario, M5S 1A7, Canada}

\author{S.~Kasahara}
\author{Y. Matsuda}
\affiliation{Department of Physics, Kyoto University, Kyoto 606-8502, Japan}

\author{T. Shibauchi}
\affiliation{Department of Advanced Materials Science, The University of Tokyo, Japan}
\affiliation{Department of Physics, Kyoto University, Kyoto 606-8502, Japan}

\author{K. D. Finkelstein}
\affiliation{Cornell High Energy Synchrotron Source, Cornell University, Ithaca, New York
14853, USA}
\author{W.~Wu}
\author{S.~R.~Julian}
\author{Young-June~Kim}
\affiliation{Department of Physics, University of Toronto, 60 St.~George St., Toronto,
Ontario, M5S 1A7, Canada}
\author{I. I. Mazin}
\affiliation{code 6390, Naval Research Laboratory, 4555 Overlook Avenue SW, Washington, DC
20375, USA}
\author{L. Boeri}
\affiliation{Institute for Theoretical and Computational Physics, TU Graz, Petersgasse 16,
8010 Graz, Austria.}\date{\today }

\maketitle

\vspace{1.7cm}


\subsection{S1.  Fe K$\beta$ emission line and the size of the local moment}

The local moment sensitivity of the Fe K$\beta$ emission line $(3p
\rightarrow 1s)$ originates from a large overlap between the $3p$ and
$3d$ orbitals. This interaction is mainly driven by the presence of a 
net magnetic moment $(\mu)$ in the $3d$ valence shell~\cite{SM_Tsutsumi1976,SM_Peng1994} 
and causes the  K$\beta$ emission line to split into a main peak K$\beta_{1,3}$ and 
a low-energy satellite K$\beta^\prime$. A schematic diagram of the Fe K$\beta$ 
emission process is  shown in Fig.~\ref{SM_fig01} inset for both non-magnetic (red) 
and magnetic (blue) Fe$^{2+}$ in the atomic limit. Filled and empty circles represent 
electrons and holes, respectively, and $\Delta E$ represents the splitting of  
K$\beta_{1,3}$ and K$\beta^\prime$. Information on the local moment of Fe can be 
extracted using the overall shape of the Fe K$\beta$ emission spectra by applying 
the  integrated absolute difference (IAD) analysis.~\cite{SM_Vanko2006}

\renewcommand{\thefigure}{S1}
\begin{figure}[htb]
\includegraphics[trim=1.0cm 0cm -1cm 0cm, width=0.9\columnwidth]{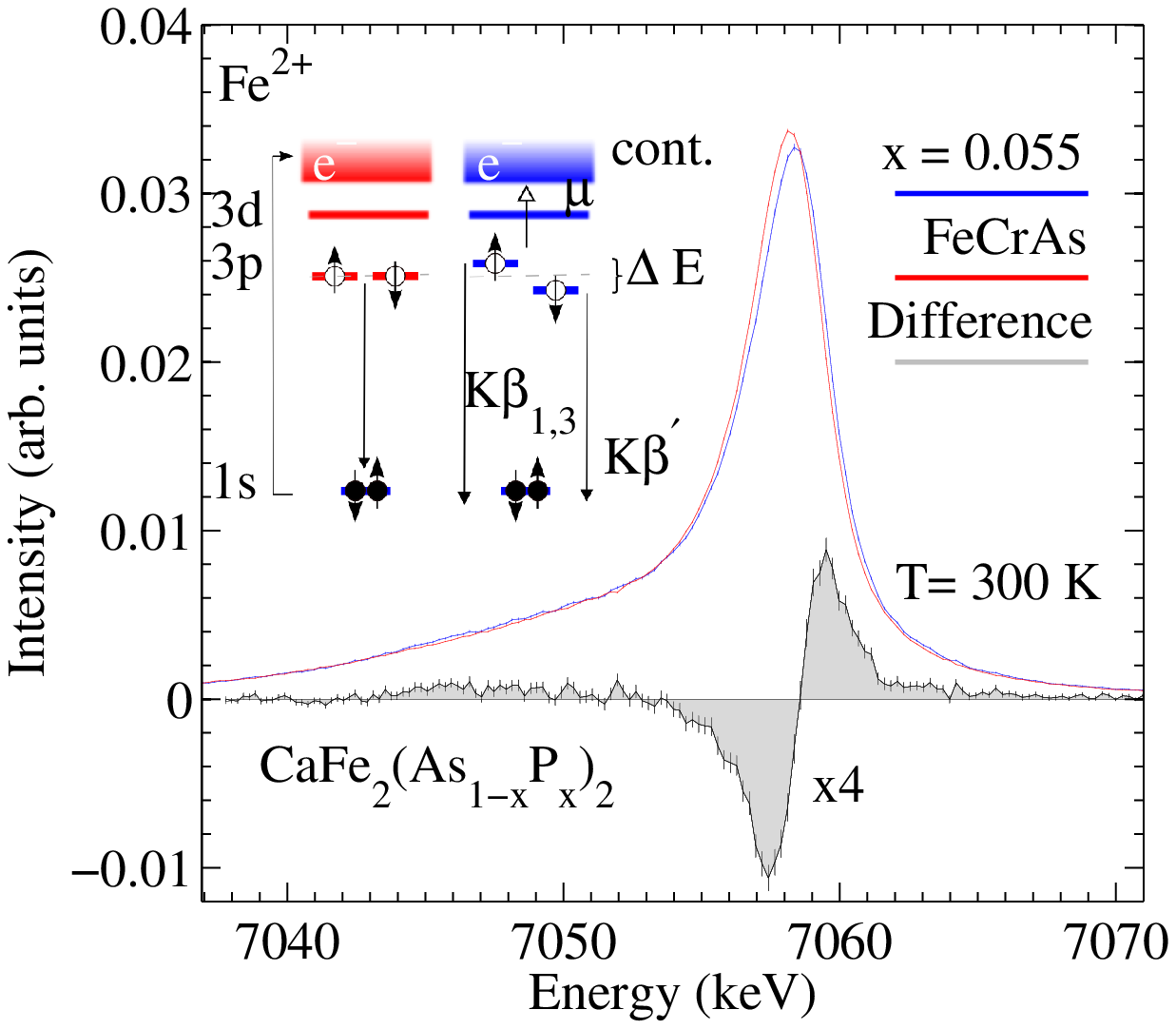}
\caption{ (Color online)  Comparison of the Fe K$\beta$ emission line (3$p$ $\rightarrow$ 1$s$) of CaFe$_2$(As$_{1-x}$P$_x)_2$ ($x = 0.055$) and FeCrAs taken at T = 300 K. Each spectrum was normalized to unity and the difference plotted (gray). The difference spectra was magnified by a factor of 4.  The inset shows the K$\beta$ emission process in the atomic limit (see text).}
\label{SM_fig01}
\end{figure}

In Fig. \ref{SM_fig01} we show the Fe K$\beta$ emission line of CaFe$_{2}$(As$_{1-x}$P$_{x})_{2}$  ($x=0.055$) 
taken at T = 300 K along with a non-magnetic FeCrAs reference spectrum.~\cite{SM_Hlynur1,SM_Wu2009,SM_Ishida1996}
Relative to the main line in FeCrAs, we see that the K$\beta_{1,3}$ peak of the $x=0.055$ sample shifts 
towards higher energy while the intensity and the width of this peak also change; a contribution from 
K$\beta^\prime$ on the lower energy side becomes visible now.  These changes are all attributed to the 
existence of a local moment. To follow the IAD procedure from Ref.~\citenum{SM_Vanko2006}, the area 
underneath each spectrum was normalized to unity. The reference spectrum was then subtracted from 
the sample spectrum, and the resulting difference plotted. For display purpose, the difference was 
magnified by a factor of 4. The IAD value can be extracted by integrating the absolute value of 
the difference spectrum. This quantity is found to be linearly proportional to the local spin 
magnetic moment of the Fe atom.~ \cite{SM_VankoPRB2006}

\renewcommand{\thefigure}{S2}
\begin{figure}[htb]
\includegraphics[trim=1.0cm 0cm 0cm 0cm, width=0.9\columnwidth]{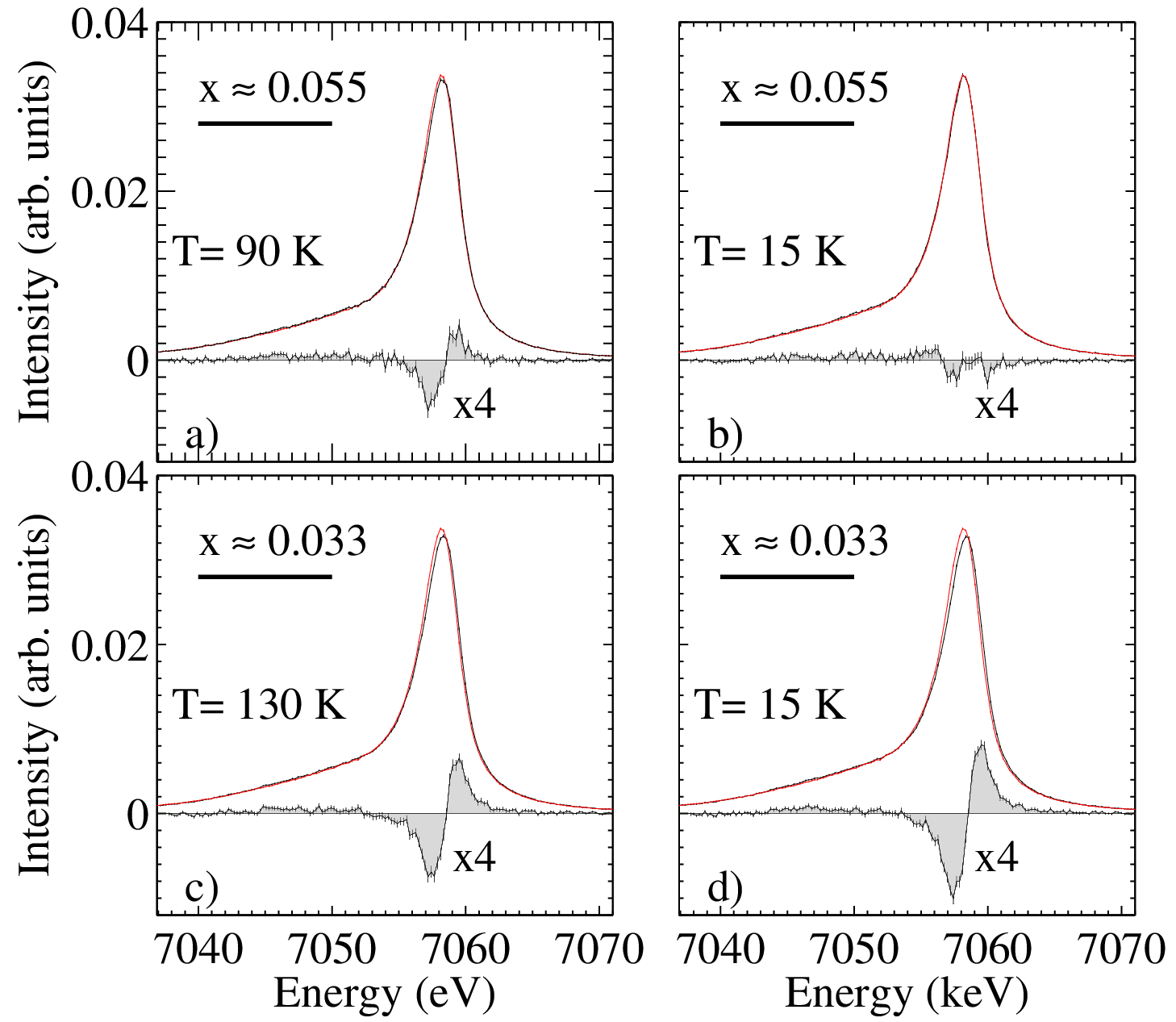}
\caption{ (Color online)  Comparison of the Fe K$\beta$ emission line (3$p$ $\rightarrow$ 1$s$) of CaFe$_2$(As$_{1-x}$P$_x)_2$ (black line) and FeCrAs (red line)   for $x = 0.055$ at (a) T = 90 K and (b) T = 15 K, and for  $x = 0.033$ at (c) T = 130 K and (d) T = 15 K. Each spectrum was normalized to unity and the difference plotted (gray). The difference spectra was magnified by a factor of 4.}
\label{SM_fig02}
\end{figure}

\subsection{S2.  Temperature dependence of the Fe K$\beta$ emission line}

Fe K$\beta$ emission lines obtained at different temperatures are shown in Fig. \ref{SM_fig02} (a) - (d) 
for both $x = 0.055$ and $x = 0.033$ samples. At T = 300 K the $x = 0.033$ sample shows the same characteristics 
as the $x = 0.055$. However, at T = 15 K significant changes can be observed, the K$\beta_{1,3}$ in the $x = 0.055$ 
sample shifts towards lower energy and the contribution from K$\beta^\prime$ is suppressed.  This is well 
captured in the difference spectra, which shows a complete suppression and provides evidence for a quenched local moment.
The $x = 0.033$ sample show opposite trend, from T = 130 K to T = 15 K the difference spectra grows modestly indicating that the local moment 
increases as the sample enters the magnetically ordered tetragonal phase.